\documentclass{elsart}
\usepackage{epsfig}
\begin{document}
\def\const{\mbox{const}}
\def\e{{\rm e}}
\def\al{\alpha}
\def\eps{\epsilon}
\def\be{\begin{equation}}
\def\ee{\end{equation}}
\def\d{\partial}
\def\l{\left(}
\def\r{\right)}
\def\la{\langle }
\def\ra{\rangle }
\def\e{{\rm e}}
\def\eps{\epsilon}
\def\d{\partial}
\def\half{\frac{1}{2}}
\newcommand{\tg}{\mathop{\rm tg}\nolimits}
\renewcommand{\ln}{\mathop{\rm ln}\nolimits}
\newcommand{\sm}[1]{{\scriptscriptstyle \rm #1}}
\newcommand{\Tr}{{\rm Tr}}
\newcommand{\bg}{\begin{gather}}
\newcommand{\eg}{\end{gather}}
\newcommand{\gtrsim}{\ge}
\newcommand{\lesssim}{\le}
\newcommand{\eq}[1]{(\ref{#1})}
\newcommand{\EP}{\mbox{e$^+$}}
\newcommand{\EM}{\mbox{e$^-$}}
\newcommand{\EPEM}{\mbox{e$^+$e$^-$}}
\newcommand{\EMEM}{\mbox{e$^-$e$^-$}}
\newcommand{\EE}{\mbox{ee}}
\newcommand{\GG}{\mbox{$\gamma\gamma$}}
\newcommand{\GP}{\mbox{$\gamma$e$^+$}}
\newcommand{\GE}{\mbox{$\gamma$e}}
\newcommand{\LGE}{\mbox{$L_{\GE}$}}
\newcommand{\LGG}{\mbox{$L_{\GG}$}}
\newcommand{\LEE}{\mbox{$L_{\EE}$}}
\newcommand{\TEV}{\mbox{TeV}}
\newcommand{\WGG}{\mbox{$W_{\gamma\gamma}$}}
\newcommand{\GEV}{\mbox{GeV}}

\begin{frontmatter}

\title{TESLA: Potentials of $\gamma\gamma$ and $e^+e^-$ Options in Stoponium
Searches}

\author[INR]{D.S.~Gorbunov,}
\author[SINP]{V.A.~Ilyin,}
\author[INP]{V.I.~Telnov}

\address[INR]{Institute for Nuclear Research, Moscow, 117312, Russia}
\address[SINP]{Institute of Nuclear Physics, Moscow St. Univ., Moscow,
  Russia} \address[INP]{Institute of Nuclear Physics, 630090,
  Novosibirsk, Russia and DESY, Hamburg, Germany}

\begin{abstract} 

In some supersymmetric extensions of the Standard Model fairly light
superpartner of t-quark is predicted, which may form bound states ({\it
stoponiums}) under certain conditions. We estimate potentials of TESLA linear
collider in search for stoponium, considering the basic electron-positron
option and the $\gamma\gamma$ option (Photon Linear Collider - PLC). 

It is found that PLC could be the best machine for discovery and study of
these new narrow strong resonances. It can produce thousands stoponiums per
100 fb$^-1$ integrated \GG\ luminosity in the high energy peak. In the case
of scenarios when stoponium decays mainly into two gluons the {\it
signal/background} ratio is about 1/4. In addition the decay channel $S\to
hh$ into two lightest Higgs bosons could be seen with high significance.
Thus, several weeks run is sufficient for the discovery of stoponium, if
its mass is approximately known (for example from observation of direct
stops production at LHC). Then, in MSSM scenarios with  dominant $S\to hh$
decay PLC shows excellent possibilities to discover bound state of stops,
practically immediately after beginning of operating. Thus, PLC has good
possibilities to study strong interaction of top quark superpartners in
nonperturbative regime.

The \EPEM\ option also has some prospects to observe stoponium but only in
the case of  scenarios with  dominant  decay into two lightest Higgs
bosons, with tens of events per 100 $fb^{-1}$. Interesting possibility
appears in the case when the resonance is seated on 0.1\% width luminosity
peak -- one could resolve the stoponium exited states. 

\end{abstract}

\end{frontmatter}
\newpage

\section{Introduction}

The broken supersymmetry is favorite among the different extensions of  the
Standard Model. It can happen that superpartners of top-quarks ({\it stops},
$\tilde{t}$) are long-living enough to compose (colorless) bound states, {\it
stoponiums}, denoted as $S$ in what follows. In this scenario experimental
study of the corresponding resonances could provide precise value of stop mass
and stoponium partial widths, consequently yielding precise values of various
parameters of SUSY Lagrangian. Then, if the difference between stop and LSP
masses is very small, the search for stop evidence in collisions at high energy
could be problematic. Observation of stoponium bound states will be the
signature of such models confirming the existence of stop. 

There are theoretical motivations for stop to be fairly light. First one
appeals to the renormalization group  behavior of soft mass terms. Indeed,
gauge couplings raise while Yukawa couplings reduce these terms when energy
scale evolves down, with Yukawa contributions being very large for stop.
The next motivation concerns left-right mixing in squark sector, which is
proportional to Yukawa coupling and decreases the mass of the lightest
stop. Therefore, light stop may appear in different SUSY models ( see,
e.g., Refs.~\cite{lightstop} for examples in the frameworks of supergravity
and gauge mediation). Experimental bound on stoponium mass comes from
searches for stop at LEP2 and TEVATRON, it depends on the
MSSM spectrum~\cite{stop-exp}: lower bound is about 90 GeV for sneutrino
masses larger than 45-50 GeV or for neutralino masses larger than 50 GeV
(ALEPH), while CDF excludes stop mass up to 130 GeV for smaller sneutrino
masses. The limitation is weaker if stop and neutralino masses are
degenerate, it is about 60 GeV (ALEPH).

Stoponium was studied in Refs.~\cite{stoponium,drees} in detail, in
particular its effective couplings and partial widths were calculated, and
prospects to be discovered at LHC were estimated. In Ref.~\cite{drees} it
was briefly mentioned also the possibility to observe stoponiums in photon
collisions, however, without analyzing this phenomenology. The  first look
at the PLC prospects on the stoponium search was done in \cite{we}. 

Now, when main parameters of TESLA project are under technical
discussion\footnote{see {\sf
http://www.desy.de/$\sim$njwalker/ecfa-desy-wg4/parameter\_list.html} for
current TESLA reference parameter set} one should understand clearly
signatures of stop bound states in $e^+e^-$ and $\gamma\gamma$ collisions,
and compare potentials of these two options of future linear collider. In
present analysis we will use year integrated luminosity for the $e^+e^-$
option equal to 300 fb$^{-1}$ for $\sqrt{s_{e^+e^-}}=500$ GeV, and one
should rescale this figure for lower collision energies approximately as
${\it L}\sim \sqrt{s_{e^+e^-}}$ \cite{brinkTDR}. Beamstrahlung and ISR will
affect considerably the stoponium production rate due to a condition to
tune the collision energy at the resonance point. According to the current
TESLA reference parameter set the average energy loss due to beamstrahlung
is about 3\%. The average energy loss due to ISR is about 5\% \cite{NLC}.
Correspondingly, the luminosity distribution has a characteristic width
about 5\%.

However, stoponium is very narrow resonance, even more narrow than the
initial beam energy spread. Its production rate is proportional to the
differential luminosity, $dL/dW$, at the $W = \sqrt{s_{ee}}$ peak, where
$W$ is a collision energy for hard subprocess. The initial (beam) energy
spread for energies 200--500 GeV is about 0.07--0.1\% and it is determined
by the bunch compression system and undulator for the positron production.
The fraction of the luminosity in this $\pm$0.1\% peak is determined by the
ISR and beamstrahlung. The ISR leaves in this peak about 50\% of the
luminosity~\cite{NLC}. The average number of beamstrahlung photons with the
energy more than $0.001\sqrt{s_{ee}}$ is about $N_{\gamma} \sim 1.15$ per
electron for TESLA conditions. Thus, the probability of the \EPEM\
collision without such beamstrahlung photons can be estimated as
$(1-e^{-N_{\gamma}})^2/N_{\gamma}^2 \sim 0.35. $ The probability of events
without any photons (ISR or beamstrahlung ones) with the energies greater
0.1\% is $0.5\times 0.35 \sim 0.17.$ So, about 17\% of the luminosity is
concentrated in $\pm$0.1\% range. The differential luminosity in this peak
is higher than in 5\% interval by a factor of $(5/0.2)\cdot 0.17/0.8 \sim
5.5$ times (here factor 0.8 is due to ISR in the 5\% region). This peak
could be very important factor for increasing significance of the resonance
(when its mass is known), for measurement of its mass and even for
resolving close excited states of the stoponium. Note, however, that for a
search for the stoponium this peak will not help, because in presence of
large background the scanning time in some wide energy interval, required
for the resonance observation, has the following dependence on the bin
width  $\Delta_W$  and the differential luminosity, $t_{scan} \propto
1/(\Delta_W(dL/dW)^2)$ (here bin $\Delta_W$ corresponds to the effective
width of the luminosity distribution in the peak). So, the ratio of
scanning times for the 5\% and the 0.1\% width peaks is
$(0.2/5)\times(5.5)^2 \sim 1.2$, almost the same.

Photon colliders based on Compton backscattering of laser photons on high
energy electrons has been proposed a long time ago~\cite{backscattering}.
This option  has been included in the TESLA Conceptual Design
Report~\cite{PLC} and work on the Technical Design Report is under way.
Since the CDR parameters of TESLA were changed and luminosities have grown
both in \EPEM\ and \GG\ collisions.

At the present workshop it was reported \cite{telnovPLC-TESLA} that PLC
luminosity can be further increased by a factor of 2.5 due to possible
decrease of the horizontal beam emittance at the TESLA damping ring
(however this has not improved \EPEM\ luminosity because it is restricted
by collision effects).  At present the \GG\ luminosity within the 20\%
interval just below $\sqrt{s_{\gamma\gamma}}=0.8 E_{ee}$ could be about
40\% of luminosity in the \EPEM\  collisions (where $L_{\EPEM} = 3\times
10^{34}$ at $\sqrt{s_{ee}}=500$ GeV). In the analysis we will consider 60
fb$^{-1}$ for PLC year luminosity at $\sqrt{s_{\gamma\gamma}}\le 400$ GeV.

The most bright evidence of the narrow resonance is its direct s-channel
production in $e^+e^-$ annihilation or $\gamma\gamma$ fusion. One can note,
that high powers of the coupling constants, $\alpha^2\alpha_s^5$, emerge in
the squared matrix elements in the $e^+e^-$ annihilation into stoponium.
Indeed, $\alpha^2$ arises from two electroweak vertices, and $\alpha_s^5$
comes from squared derivative of the stoponium wave function (scalar
stoponium can be created there only in P-wave  by propagation of  neutral 
vector particle: photon or $Z$ boson). At the same time two powers of
$\alpha_s$ are eliminated  in the case of $\gamma\gamma$ fusion mechanism
because the stoponium production can be proceeded in S-wave there. This
circumstance makes for the relative enhancement of the stoponium production
rate at PLC in comparison with $e^+e^-$ option. Two powers of $\alpha_s$
are eliminated also in the case of associated production of stoponiums, for
example in the {\it Higgs-like} reactions $e^+e^-\to ZS$ and 
$e^+e^-\to\nu\bar{\nu}S$. Hereafter we denote stoponium as $S$.

To complete this brief review of possible production mechanisms one can note
that in hadron collisions the stoponium resonance production is available  in
S-wave through the gluon fusion. Here the effective $ggS$ vertex includes
$\alpha_s^{5/2}$ and one can anticipate large stoponium cross sections.
However, main decay channel $gg$ is too dirty due to huge QCD $2jets$
background. Then, the most promising decay channel at LHC is $\gamma\gamma$
\cite{drees}, but in order to discover stoponiums one year of LHC operating at
high luminosity is needed, or even more depending on SUSY scenario. 

We consider stoponium mass range $M_S=200-400$~GeV, which could be surely 
probed by {\it first} TESLA run. It is worth to note that the same interval
is not an exceptional case for SUSY models with stoponiums as a
quasistationary state, as we discuss briefly in the next section. 

Cross sections for various tree-level background processes were evaluated with
the help of CompHEP package \cite{CompHEP}.

\section{Stop bound states}

It is clear that gluons try to bind two stops as well as ordinary quarks. The
corresponding bound state can be described as a quasistationary system with
energy levels $E_n$ ($<0$) and masses $M_n=2m_{\tilde{t}}+E_n$ similarly to
quarkonium. For stoponium mass $M_S=200-600$~GeV the binding energies $E_n$ are
of order 1 GeV~\cite{ng}. This treatment is valid if the formation process
(time scale $\sim |E_n|^{-1}$) is faster than destroying one.  

Among destroying mechanisms the obvious ones are the stop
decays\footnote{We suppose R-parity to be conserved, as favored by the
absence of rapid proton decay and lepton flavor violating processes.}:
$\tilde t\to t+$LSP, $b+chargino$ and $c+neutralino$. At first, let us
consider the third decay. It proceeds only through loop diagrams,  if
Universal boundary condition on soft terms is imposed (that is motivated by
the absence of FCNC). So, partial width is  highly reduced  by a factor of
$\sim 10^{-7}$ in comparison with the first two tree-level decay
processes~\cite{c-neutralino}. The rates of latter decays depend on the
parameters of the model. As an example, in the framework of gravity
mediation, where LSP is neutralino, these decays proceed at the tree level
and the corresponding partial widths are of order {\it O}$(\alpha m_{\tilde
t})$. In the framework of models with gauge mediated supersymmetry breaking
\cite{revGMM}, where LSP is gravitino, the first process is strongly
suppressed by  supersymmetry breaking scale, but remaining one has the same
partial width as in gravity mediation. Hence, the possibility of existence
of stoponium is a subject of special study in each concrete model. For
instance, in models with the lightest chargino being mostly wino and the
lightest stop being mostly right stop (i.e., $m_{t_L}>m_{t_R}$), decay into
chargino is damped and stoponium could exist if $m_{\tilde t}-m_{LSP}<m_t$,
i.e., when the first decay channel is kinematically forbidden. One can
state that SUSY scenario, where tree-level decays, $\tilde t\to t+$LSP  and
$\tilde t\to b+chargino$, are somehow suppressed and, therefore, stop decay
can not destroy the stoponium formation, is not an exceptional case.

Next destroying mechanism is related to the stop annihilation. Here two gluon
channel is always open with partial width  about 1 MeV. Generally the gluon
channel is dominant. However, for the certain choice of model parameters,
partial width into two lightest Higgs bosons, $S\to hh$, can be larger,
increasing the stoponium total width by a factor of $\sim 5-10$. In
Ref.~\cite{drees} these figures were analyzed and found that quasistationary
description is valid for $M_S<600$~GeV in models with forbidden stop tree-level
decays and neutralino being mostly bino. The worst case is a model with
chargino and neutralino states are both higgsino-like. Here the stoponium total
width increases rapidly with $M_S$ and quasistationary treatment fails for
$M_S>300$~GeV.

\section{Stoponium in $\gamma\gamma$ collisions}

Let us begin with study of stoponium events in photon-photon scattering. The
main effect associated with stoponium would be a direct resonance
production, where stoponium is produced in spin 0 state.
The corresponding  cross section is described by the 
Breit-Wigner formula (similarly to the light Higgs
production discussed in Ref.~\cite{Borden})  
\begin{equation}
\sigma_{\gamma\gamma\to S\to f}(\hat{s})= 8\pi 
\frac{\Gamma_{\gamma\gamma}\,\Gamma_f}{\l \hat{s}-M_S^2\r^2+ \Gamma_{\rm
tot}^2M_S^2}(1+\lambda_1\lambda_2) \;, 
\end{equation}
\noindent
where $\hat{s}=W^2$ is squared colision energy for hard subprocess,
$\Gamma_f$ is the stoponium partial width for the decay into
state $f$, $\Gamma_{\rm tot}$ is stoponium total width,
$\lambda_{1,2}$ are helicities of initial photons.

  At photon colliders the width of the luminosity distribution is much wider
than that of the stoponium. After integration over the luminosity distribution
we obtain the effective cross section
\begin{equation}
\sigma_f  = \frac{1}{L_{\GG\ }} \frac{dL_{\GG\ }}{dW}  
\frac{4\pi^2 \Gamma_{\GG\ } Br_f}{M_S^2}\;
(1+\lambda_1\lambda_2),
\end{equation}
The differential luminosity, $dL_{\GG\ }/dW$, at $W=M_S$ can be
estimated as $L_{\GG\ }/0.15M_S$ according to 15\% width of the high
energy luminosity peak (note, we define the \GG\ luminosity as the
luminosity in the interval $\Delta W/W_{max} = 20$\%). So the effective
cross section is
\begin{equation}
\sigma_f = \frac{(4/0.15)\pi^2 \Gamma_{\GG\ } Br_f}{M_S^3}\; 
(1+\lambda_1\lambda_2)\,.
\end{equation}
Photon beams are planned to be highly polarized. Hence, as stoponium is a
scalar the production cross section will be enhanced by factor two if
initial photons have total helicity equal to zero. Hereafter we assume
initial total helicity 0 ($\lambda_1\lambda_2=1$) in numerical estimates.
Finally, we parameterize the stoponium cross sections as follows
\begin{equation}
\sigma_f \;\approx\; 50\mbox{fb}\cdot (1+\lambda_1\lambda_2)\cdot 
   \left(\frac{Br_{\gamma\gamma}}{4\cdot 10^{-3}}\right) \cdot
   Br_f \cdot
   \left(\frac{\Gamma_{tot}}{\mbox{1MeV}}\right) \cdot
   \left(\frac{200\mbox{GeV}}{M_S}\right)^3 .
\label{gamgam}
\end{equation}
\noindent
One should take into account that squared stoponium wave function at
the origin, attending in $\Gamma_{tot}$, scales as a square root of
its mass~\cite{ng}.

As it has been stressed above one can discuss two main variants of the SUSY
models, one with dominant $gg$ decay mode and another with stoponium total
width being saturated by $hh$ mode. Let us make qualitative {\it
signal/background} analysis for different decay channels within these two
variants. The signal significance can be evaluated by ratio
$N_S/\sqrt{N_B}$ because one deals with resonance and background rate in
the signal bin can be fixed as average cross section in neighboring bins
(here $N_{S,B}$ are numbers of signal and background events). We used the
results for stoponium width and branching ratios calculated in
Ref.~\cite{drees} with some corrections~\cite{we}.

\vspace{0.2cm}    
\noindent
{\bf 1.} 
In the first scenario stoponium total width $\approx 1.3$~MeV  and
photon-photon branching is ${\rm Br}_{\gamma\gamma}\approx 3.4\cdot 10^{-3}$.
By making use of Eq.(\ref{gamgam}) one obtains the signal rate at the level
of 110 fb for $M_S=200$ GeV. So, more than six thousand stoponiums will
be produced per year if {\it L}$_{PLC}^{year}=60$ fb$^{-1}$. Background is
two jet production, where subprocess $\gamma\gamma\to q\bar q$ gives main
contribution with very large unpolarized cross section, $\sim 6$ pb, if
optimal cut on the jet angle of 45$^\circ$ is applied. However, production
of fermions is suppressed in collisions of photons with the total helicity
0~\cite{Ispirian}:
\begin{equation}
d\sigma(\GG\ \to q\bar{q})/d cos{\theta} \propto \frac{1+cos^2{\theta}}
{1-cos^2{\theta}}(1-\lambda_1\lambda_2) 
\qquad \mbox{for}\; \beta \longrightarrow 1\;,
\end{equation}
where $\beta$ is velocity of the quarks. So, $q\bar{q}$ pairs are produced
only in collisions of photons with the total helicity 2. This 
fact\footnote{Note, that in \cite{we} this background was overestimated by
taking the unpolarized cross section.} is used for suppression of similar
background in the analysis of the Higgs production in \GG\ collisions
\cite{Barklow,Borden,Soldner}. The corresponding luminosity spectra for
total helicity 0 ($L_0$) and 2 ($L_2$) can be found elsewhere
\cite{telnovPLC-TESLA}. In the region $\sim$10\% near the maximum energy
(detector resolution or intrinsic resolution of the PLC for two collinear
jets) one can obtain the ratio of the luminosities $L_2/L_0 < 0.1$, that
assumes at least one order suppression of the $q\bar{q}$ background.  Note
that the cross section 6 pb for background corresponds to the case of
unpolarized beams. It is zero for collisions of photons with the total
helicity 0 and is equal to 12 pb for the total helicity 2, therefore the
remaining cross section is 1.2 pb. Note that due to gluon emission
$q\bar{q}$ pairs can be produced even in collisions of photons with the
total helicity 0. Detailed studies have shown that with proper cuts this
process is not important \cite{NLO,resolved} and contribution from resolved
photons is also not significant \cite{resolved}.

Furthermore, the cross section is proportional to the fourth power of
the electric charge of quarks, so the main contribution is given by
$u$ and $c$ quarks. The later can be easily suppressed by the vertex
detector. This gives additional factor of 2 in the background
suppression. Additional improvement can give the detector energy
resolution which is at least factor of 2 smaller than the width of the
\GG\ luminosity peak. All three methods give a suppression factor of
40. The remaining background is about 0.3 pb, while the isotropic
signal is smaller by factor 1.4 only if the cut on the jet angle of
45$^\circ$ is applied. Hence in the scenario with dominant $gg$
channel the {\it signal/background} ratio is $\sim 1/4$.  The signal
significance is about 35 for 60 fb$^{-1}$ and $M_S=200$~GeV. 

These figures can be related to the first year of PLC operation if the
stoponium mass is known approximately, for example from the
observation of the direct stops production at LHC. In opposite case,
for a search of a stoponium one should make scanning with the energy
bin $\Delta_W/W \sim$ 10\%.  It is clear that stoponium can be found
in this scenario during several month work in the whole energy region
under discussion, 200-400 GeV.

For two photon channel the background process,
$\gamma\gamma\to\gamma\gamma$, proceeds through one-loop diagrams, so the
corresponding cross section is small, about 10 fb~\cite{JT1}. One should
note that the photon-photon invariant mass bin can be taken equal to
$2\mbox{GeV}\cdot \sqrt{M_S/100\mbox{GeV}}$ for CMS-like crystal
electromagnetic calorimeter \cite{CMS}. Thus, for $M_{\gamma\gamma}=200\pm
1.4$ GeV window the background rate can be estimated at the level of 1 fb.
The signal rate is $0.4\div 0.14$ fb for $M_S=200\div 300$ GeV, providing
the signal significance about $3\div 1.1$  for statistics 60 fb$^{-1}$.

Some other decay channels in the framework of the first scenario should be
discussed. First note, that $WW$ final state has no chance for the detection of
stoponiums due to huge SM background, $\sigma^{tot}_{\gamma\gamma\to WW}\sim
60$ pb at $\sqrt{s_{\gamma\gamma}}=200$ GeV. More promising are decay channels
with background processes emerging due to the higher order corrections from
perturbation theory. For instance, SM background to $\gamma Z $ and $ZZ$ final
states comes from 1) one-loop $\alpha^4$ processes $\gamma\gamma\to\gamma Z$
($10-15$ fb~\cite{JT2}) and $\gamma\gamma\to ZZ$ ($\sim 50$ fb~\cite{ZZ}), and
2) from tree-level $\alpha^3$ processes (e.g. $\gamma\gamma\to\gamma q\bar q$
for $S\to \gamma Z\to\gamma +2jets$), with total cross section smaller than 1
fb within cuts on final $\gamma$ and jets reasonably motivated by 2-body
($\gamma+Z$) kinematics of the signal events.

As to signal $\gamma Z$ rate one can get from Ref.~\cite{drees} the branching 
${\rm Br}_{\gamma Z} \sim 2\cdot10^{-3}$, so $\sigma_{\gamma Z}\sim 0.22$ fb
already for $M_S=200$ GeV. It means very low level of the signal significance,
lower than 0.5 for statistics 60 fb$^{-1}$.

Natural level of $ZZ$ branching is about $4\cdot10^{-2}$ for stoponium masses
far from the threshold, 250-400 GeV, although in some points it could fall down
due to opening of new channels or degeneration of stoponium and Higgs masses.
This provides signal rate  $\sim 2.5$ fb for $M_S=250$ GeV and significance at
the level of 2.7 if one uses formula (\ref{gamgam}). However, the threshold
effect is significant still for this value of the stoponium mass, and these
figures should be improved to 1.8 fb for signal rate  and to 2 for the
significance.

The $hh$ decay channel, where $h$ is the lightest Higgs boson, is open if
$M_S>2m_h$. As current limit on $h$ mass is about 80-100 GeV this channel
could exist for $M_S>200$ GeV. If consider mass region  not very close to
the threshold  (say $M_S>300$ GeV for $m_h=115$ GeV)  the $hh$ branching is
about $2\cdot10^{-2}$ or even higher. In this case the signal rate is about
1~fb or larger, if one uses formula (\ref{gamgam}), and if take into
account the threshold factor one gets signal cross section at the level of
0.2 fb or larger. The background from direct double $hh$ production through
one-loop diagrams can be estimated by the cross section of this process in
SM, $\sim 0.2$ fb \cite{hh}. There are no reasons for very large additional
contributions to this process in supersymmetric models. Then, we found that
direct electroweak production of four b quarks ($\gamma\gamma\to bbbb$)
together with contribution from $cccc$ final state (assuming 10\% of $b/c$
misidentification) has the rate smaller than 0.1 fb. These cross sections
for background processes correspond to 15\% width of the \GG\ luminosity
spectrum. The detector resolution is at least factor of two better (full
width), therefore the total cross section of background processes can be
estimated as (0.2+0.1)/2 = 0.15 fb. We see that $hh$ decay can also be
studied in the considered scenario. For $M_S=300$~GeV, $m_h=115$ GeV and 60
fb$^{-1}$ integrated luminosity about 20 stoponium events will be produced
in the decay channel $S\to hh$ with $S/B$ ratio about 2.3 and statistical
significance about 7.

\begin{figure}[htb]
\begin{center}
{\epsfig{file=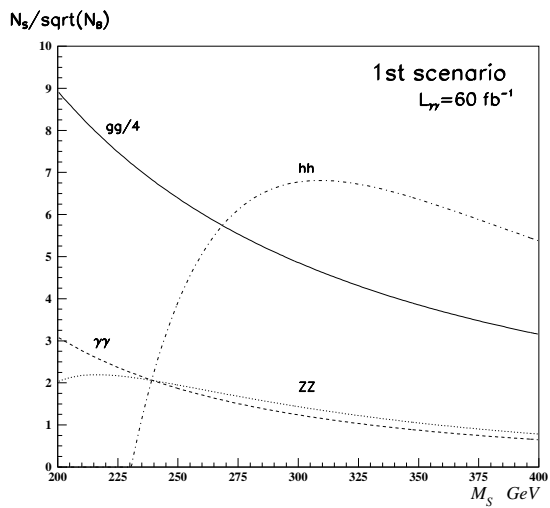,width=8cm}}
\end{center}
\caption{Signal significance of stoponium events in various channels for
the first scenario; mass of the lightest Higgs boson is taken equal to
115~GeV.}
\label{gg-fig}
\end{figure}

In Fig.~\ref{gg-fig} the signal significances are represented for main
stoponium decay channels in the case of the first scenario. 
As a resume for this scenario we conclude that stoponium can be found
at PLC during several months scan of the 200--400 GeV region in $gg$ and
$hh$ decay modes. If its mass is known approximately, it will be found
during first weeks. More than a year is necessary in order to observe
stoponium in $\gamma\gamma$ and $ZZ$ channels.

Note that LHC (at the high luminosity operating stage) has good prospects
to observe light stoponium in $\gamma\gamma$ mode in this
scenario~\cite{drees}. Thus, these two colliders could be complementary in
study of different effective stoponium couplings, $S\gamma\gamma$ and $Sgg$
at LHC and the photon collider,  respectively  and $Shh$ at PLC if this
decay channel is open kinematically.

\begin{figure}[htb]
\begin{center}
{\epsfig{file=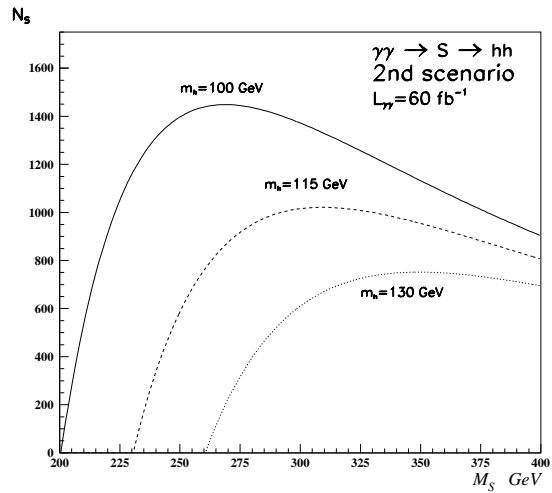,width=8cm}}
\end{center}
\caption{Total number of stoponium events in dominant $hh$-channel for
the second scenario.
Background (direct production of two light Higgses and $4b-jets$)
is expected less than 100 events.}
\label{hh-fig}
\end{figure}

\vspace{0.2cm}   
\noindent   
{\bf 2.}   
In the second scenario stoponium total width could be about 10 MeV or even
larger. The photon-photon branching in this case is smaller, $\sim
(2-4)\cdot 10^{-4}$. So, about thousand of stoponiums will be produced per
year and almost all of them will decay to pairs of lightest Higgs bosons.
This result suggests, that stoponium will be discovered practically
immediately after PLC start, since the background ($hh$, $bbbb$ ...) is
very small. Again the scanning over the  energy interval is necessary if
the stoponium mass is not known. In Fig.~\ref{hh-fig} the year yields of
stoponiums are represented for different masses of lightest Higgs bosons.

Due to rather high statistics for the signal and absence (practically) of
the background one gets PLC as a {\it stoponium factory}. The detailed
study of the stoponium characteristics will be available in this case, in
particular measurement of its mass and total width and effective couplings
$Shh$ and $S\gamma\gamma$. One can stress the importance of study the $Shh$
coupling, which relates directly to the {\it stop-Higgs} interaction and,
thus, to the mechanism of the superpartner mass generation.

Note that in this scenario several years of operating at high luminosity is
needed in order to observe stoponium at LHC \cite{drees}.

\section{Stoponium in $e^+e^-$ collisions}

{\bf 1.} 
First we discuss direct resonant production of stoponiums in $e^+e^-$
collisions where initial electron and positron would be in P-state
(thus, stoponium is produced in spin 1 state). 
As in case of $\gamma\gamma$ collisions one gets for narrow resonance
after the integration of Breit-Wigner distribution the following formula for
production cross section
\begin{equation}
\sigma_S =
\frac{1}{L_{e^+e^-}}\frac{dL_{e^+e^-}}{dW}
\cdot \frac{6\pi^2\Gamma_{e^+e^-}}{M_S^2}
\cdot (1-\lambda_-\lambda_+)\;,
\end{equation}
where $\lambda_\mp$ are helicities of initial electrons and positrons.
The differential luminosity at $W=M_S$ can be estimated as
$L_{e^+e^-}^{\Delta_W}/\Delta_W$, where $L_{e^+e^-}^{\Delta_W}$ is a fraction
of the $e^+e^-$ luminosity corresponding to some interval $\Delta_W$ near the
$W_{max}=\sqrt{s_{e^+e^-}}$ peak. Let us consider two methods of the
stoponium detection:

1) $L_{\Delta_W}=80$\% $L_{\EPEM}$, $\Delta_W/W \sim 5$\%;

2) $L_{\Delta_W}=15$\% $L_{\EPEM}$, $\Delta_W/W \sim 0.2$\%. 

Then, the stoponium partial width into $e^+e^-$ pair, $\Gamma_{e^+e^-}$,
is equal to~\cite{nappi} 
\begin{equation}
\Gamma(S\to e^+e^-)=\frac{32}{3}\cdot \alpha^2\cdot \frac{|R_P'(0)|^2}{M_S^4}\;.
\end{equation}
with $R_P(\vec{r})$ being wave function of P-state. In correspondence with
Ref.~\cite{ng} one can approximate 
$\frac{|R_P'(0)|^2}{M_S^4}=5\cdot10^{-6}$~GeV.  

We assume, also, 80\% circular polarization for the electron beam and 60\%
for the positron beam, and neglect fairly weak enhancement by  $Z$-boson
pole for light stoponiums. 

Finally, for the first method of the stoponium detection
one gets for the production rate in $e^+e^-$ collisions the following estimate 
\begin{equation}
\sigma_S \sim 0.2\mbox{fb}\cdot (200~{\rm GeV}/M_S)^3 \,,
\end{equation}
while it is about 5 times larger in case of second method.

Let us now consider background for the direct resonance production of
stoponiums in $e^+e^-$ collisions. One should note that stoponium, being
produced in excited state with spin 1, will be transfered to the basic
state with spin 0 by the emission of a photon. Therefore, one can consider
decay modes of scalar stoponium in case of $e^+e^-$ production as well.
Then, additional photon emmited by excited stoponium could be used for
further suppression of the background. However, the detection of this
photon could be not easy because its energy is rather small, of order 1
GeV.

In the case of first MSSM scenario with $gg$ decay channel being dominant
the main background is $e^+e^-\to 2jets$ process. It has a cross section
about 10 pb for $\sqrt{s}=200$ GeV. So, this channel is too dirty with
negligible significance.

In the case of second MSSM scenario with $hh$ dominant decay mode almost
all stoponiums will decay into pair of lightest Higgs bosons, and the
number of events for $M_S =300$ GeV and 100 $fb^{-1}$ is 5 (25) (the second
number is for the case of seating on the 0.1\% luminosity peak).  As direct
pair production of Higgs bosons, $e^+e^- \to hh$, is negligible since the
corresponding $eeh$ coupling includes electron mass, this channel is should
be practically free of background. Indeed, direct production of four
b-quarks in electron-positron collisions has very small cross section if
exclude $Z$ peaks (less than 0.05fb if apply the cut $M_{b\bar b}>95$ GeV).
So, stoponium can be observed at $e^+e^-$ machine if the collision energy
is tuned at $M_S$. 

Note, that in this scenario the yield of stoponiums at PLC with 
$L_{\GG}/L_{\EPEM}=0.4$ is higher than that in \EPEM\ collisions by a
factor of 250(50) (second number for the case of seating on the 0.1\% peak of
the $e^+e^-$ luminosity). Thus,  the scanning time in \EPEM\ mode will be
about factor of 500 longer than at PLC due to smaller cross section and
smaller energy bin (see also discussion on the scanning time in the
Introduction).

{\bf 2.} 
The second effect related to stoponium at $e^+e^-$ collider  would be a
production in Higgs-like channels. In present analysis we neglect
Higgs-stoponium mixing as well as Higgs influence on stoponium-involved
processes. First, let us evaluate effective  coupling constants between
stoponium and weak bosons, 
\begin{equation}
\mbox{\it L}=\lambda_{SZZ} SZ_{\mu\nu}Z^{\mu\nu}
           + \lambda_{SWW} SW_{\mu\nu}W^{\mu\nu}\,,
\end{equation}
\begin{equation}
\lambda_{SZZ}=\frac{0.152}{M_S^{5/4}}\cdot {\sf M}_{SZZ}
\cdot\l1-\frac{4M_Z^2}{M_S^2}+
6\frac{M_Z^4}{M_S^4}\r^{-1/2},
\end{equation}
\begin{equation}
\lambda_{SWW}=\frac{0.152}{M_S^{5/4}}\cdot {\sf M}_{SWW}
\cdot \l1-\frac{4M_W^2}{M_S^2}+
6\frac{M_W^4}{M_S^4}\r^{-1/2},
\end{equation}
where we used stoponium wave function evaluated at the origin  in
Ref.~\cite{ng};   ${\sf M}_{SZZ}$ and ${\sf M}_{SWW}$ are corresponding 
amplitudes of $\tilde{t}\tilde{t}^*\to ZZ(W^+W^-)$ processes evaluated in
Ref.~\cite{drees}. Certainly, these effective couplings are obtained with
all particles being on-shell, that is rather rough approximation. However,
one can hope that it is acceptable for the estimate.

The calculation with the effective couplings above gives the cross section
about 0.03 fb for $e^+e^- \to ZS$ (in case of no mixing in stop sector and
$M_{\tilde{b}}=1$~TeV). So, only a few signal events will be produced for
100 fb$^{-1}$ statistics. The $Zjj$ background is too heavy for this level
of the signal, so only the second scenario could have some prospects. The
main background will come from associated double Higgs bosons production,
$e^+e^- \to Zhh$, the corresponding cross section is of order $0.2\div 0.5$
fb for $\tan{\beta}=3\div 50$ \cite{LChh}. It gives the signal significance
less than 1.

In the case of $W$-fusion  the signal cross section is very small, less than
$10^{-3}$ fb, that closes this channel.

\section{Conclusions}

In two considered scenarios of supersymmetric extension of the Standard
Model (1st one with  dominant  $gg$ decay mode, and $hh$ being dominant in
2nd scenario) Photon Linear Collider will be the best machine to discover
and study bound state of stops, if it exists.

In case of 1st scenario stoponium will be observed at PLC in the $gg$ and
$hh$ (later if permitted kinematically) in the beginning of the
operation -- several months for scanning of whole energy region or several
weeks if the stoponium mass is approximately known.

In case of 2nd scenario, about thousand of stoponiums will be produced free of
background. It means that stoponium can be discovered at PLC practically
immediately.

Study of the effective stoponium couplings with photons,
gluons and lightest Higgs bosons will be available at PLC, latter two 
depending on the MSSM scenario. Measurement of the stoponium mass
and total width (extracted from the measured signal rate) will be possible
also.

Stoponium discovery mass range will be limited only by attainable values of
the $\gamma\gamma$ collision energy, which is discussed up to $0.8\cdot
500$ GeV. The tuning of PLC collision energy at the resonance point is
necessary within the 15\% window. These estimates were done for the case of
PLC year luminosity being 60 fb$^{-1}$.

In $e^+e^-$ collisions stoponium could be observed only in the
scenario with $hh$ decay channel being dominant, but with the rate lower
than in the PLC case by a factor of 250 (50) (the second number for seating
on the 0.1\% luminosity peak). In this comparison it was assumed that \GG\
luminosity in the high energy peak is equal 40\% of \EPEM\ luminosity. 
Search time here is about 500 times longer than in \GG\ collisions. However,
there is one important advantage of \EPEM\ collisions: by use of very
monochromatic part of the luminosity spectrum (0.1\%) one can make precise
measurement of the stoponium mass and resolve its excited states. Although 
this is possible only in the scenario when $hh$ decay dominates.

A few further comments can be made. The first one is related to the
circumstance that at photon colliders ground state of stoponium could
not be distinguished from excited states due to the detector
resolutions. Therefore, the resonance peak will include contributions
from ground state and all excitations, leading to enhancement factor
of about 2 in all cross sections~\cite{drees}. At the same time there
is a big uncertainty because of poor understanding of the stoponium
wave function, that results in 30-50\% error when the stoponium rates
are estimated~\cite{drees}.

Then, stoponium has the same quantum numbers as neutral Higgs bosons. Thus,
interesting phenomena could appear due to the interference of stoponium with
Higgs sector. This point was discussed briefly in Ref.~\cite{we}.

\vspace{0.2mm}

The authors are indebted to I.F.~Ginzburg, H.J.~Schreiber  and
other participants of the present Workshop for useful discussions. The work
of D.G. was supported in part under RFBR  grant 99-01-18410, CRDF grant
(award RP1-2103), Swiss Science Foundation grant 7SUPJ062239, Russian
Academy of Science JRP grant \# 37 and by ISSEP fellowship. The work of
V.I. has been partially supported by the CERN-INTAS 99-377 and RFBR-DFG
99-02-04011 grants.

\def\ijmp#1#2#3{{\em Int. Jour. Mod. Phys. }{\bf #1~} (19#2) #3}
\def\pl#1#2#3{{\em Phys. Lett. }{\bf B#1~} (19#2) #3}
\def\zp#1#2#3{{\em Z. Phys. }{\bf C#1~} (19#2) #3}
\def\prl#1#2#3{{\em Phys. Rev. Lett. }{\bf #1~} (19#2) #3}
\def\rmp#1#2#3{{\em Rev. Mod. Phys. }{\bf #1~} (19#2) #3}
\def\prep#1#2#3{{\em Phys. Rep. }{\bf #1~} (19#2) #3}
\def\pr#1#2#3{{\em Phys. Rev. }{\bf D#1~} (19#2) #3}
\def\np#1#2#3{{\em Nucl. Phys. }{\bf B#1~} (19#2) #3}
\def\mpl#1#2#3{{\em Mod. Phys. Lett. }{\bf #1~} (19#2) #3}
\def\arnps#1#2#3{{\em Annu. Rev. Nucl. Part. Sci. }{\bf #1~} (19#2) #3}
\def\sjnp#1#2#3{{\em Sov. J. Nucl. Phys. }{\bf #1~} (19#2) #3}
\def\jetp#1#2#3{{\em JETP Lett. }{\bf #1~} (19#2) #3}
\def\app#1#2#3{{\em Acta Phys. Polon. }{\bf #1~} (19#2) #3}
\def\rnc#1#2#3{{\em Riv. Nuovo Cim. }{\bf #1~} (19#2) #3}
\def\ap#1#2#3{{\em Ann. Phys. }{\bf #1~} (19#2) #3}
\def\ptp#1#2#3{{\em Prog. Theor. Phys. }{\bf #1~} (19#2) #3}
\def\spu#1#2#3{{\em Sov. Phys. Usp.}{\bf #1~} (19#2) #3}
\def\apj#1#2#3{{\em Ap. J.}{\bf #1~} (19#2) #3}
\def\epj#1#2#3{{\em Eur.\ Phys.\ J. }{\bf C#1~} (19#2) #3}
\def\pu#1#2#3{{\em Phys.-Usp. }{\bf #1~} (19#2) #3}

\end{document}